\documentstyle[prd,aps,epsfig,psfig,floats]{revtex}

\newcommand\be{\begin{equation}}
\newcommand\ee{\end{equation}}
\newcommand\bea{\begin{eqnarray}}
\newcommand\eea{\end{eqnarray}}
\newcommand\ba{\begin{array}}
\newcommand\ea{\end{array}}

\begin{document}
\draft
\input epsf

\twocolumn[\hsize\textwidth\columnwidth\hsize\csname
@twocolumnfalse\endcsname

\title{Comment on: Neutrino Mass Anarchy}
\author{M. Hirsch}
\address{\phantom{ll}}
\address{{Department of Physics and Astronomy, University of Southampton,
Southampton, SO17 1BJ, U.K. }}
\maketitle
\begin{abstract}
Recently Hall, Murayama and Weiner (Phys. Rev. Lett. 84 (2000) 2572)
claimed that neutrino oscillation data are well accounted for by a 
neutrino mass matrix which appears to have random entries. Here this 
claim is disputed by constructing a specific counter example. 
Structure in the neutrino mass matrix is clearly preferred over 
random matrices. 
\end{abstract}
\vskip1pc]

Current data from neutrino oscillation experiments show definite 
hints for non-zero neutrino masses and mixings \cite{nu2000}. Perhaps 
not surprisingly, these results have triggered a flood of theoretical 
attempts to construct models for the neutrino mass matrix, after all 
they provide the first window into physics beyond the standard model. 
The majority of models \cite{SelReview} are based on the assumption 
that some high-energy scale theory creates a small expansion parameter 
$\epsilon$ and the neutrino mass matrix then takes a structure of the 
form $M_0+\epsilon M_1 + ...$, where there are possibly many zeroes 
in each order of the expansion. The task in understanding the physics 
underlying the neutrino mass matrix would then be to search for 
models which are able to produce the correct texture zeroes.

The authors of \cite{HMW} claim that such attempts might be completely 
unnecessary, since current (experimental) knowledge about the neutrino 
mass matrix can be well accounted for by a mass matrix which appears 
to have entries completely chosen at random, at least if the large 
angle MSW solution to the solar neutrino problem is correct.

To demonstrate the validity of their claim the authors of \cite{HMW} 
considered the following quantities, 
\bea
R\equiv |\Delta m_{12}^2|/|\Delta m_{23}^2| \\
s_{C}\equiv 4|U_{e3}|^2(1-|U_{e3}|^2)\\
s_{atm}\equiv 4|U_{{\mu}3}|^2(1-|U_{{\mu}3}|^2)\\
s_{\odot}\equiv 4|U_{e2}|^2|U_{e1}|^2
\eea
They then constructed three types of mass matrices, namely Dirac-like, 
Majorana-like and seesaw-like. For each they ran sample calculations 
with one million matrices with coefficients taken randomly 
in the interval $[-1,1]$. Applying cuts - motivated by the neutrino 
oscillation data - on the above quantities, 
namely, $R<1/10$, $s_C <0.15$, $s_{atm}>0.5$ and $s_{\odot}>0.5$ and 
checking how many matrices pass the cuts, they conclude that mass 
matrices which do not show any structure are sufficient to account 
for all neutrino oscillation data, thus the scenario has been called 
neutrino mass anarchy.

In the following only the Majorana case will be considered. (Very similar 
arguments can be applied to the Dirac and seesaw matrices.) 
Consider the following (Majorana) mass matrix.

\begin{equation}
m_{LL} \equiv m_{22}
\left( \begin{array}{ccc}
\alpha & r_{12} & - r_{12} + \beta \\
\cdot    & 1      &  1 + \gamma \\
\cdot    & \cdot &  1 + \delta \\
\end{array}
\right)
\label{Mbest}
\end{equation}
Eq. \ref{Mbest} contains 6 parameters and thus without specifying the 
range of $m_{22}$, $\alpha$, $r_{12}$, $\beta$, $\gamma$ and $\delta$ 
is just one possibility to write down the most general (real) Majorana 
mass matrix. 

However, once we specify a range for the coefficients, structure is 
introduced ad hoc. We can then use the same procedure as in \cite{HMW}, 
i.e. take the coefficients in a given range and run a large sample of 
randomly chosen matrices. 

Choosing all 6 unknowns within $[-1,1]$ and running a calculation with 
$10^8$ randomly generated matrices, 8 \% of these pass 
all cuts, to be compared to $1$ \% in the case of neutrino anarchy. 
Choosing, motivated by typical texture models, $m_{22}=r_{12}=1$, and 
all other parameters to be of order ${\cal O}(\lambda)$, 
with $\lambda=0.22$, 66 \% of all matrices pass the cuts. 
Choosing $m_{22}=1$ and all other parameters ${\cal O}(\lambda)$, 75 \% 
of the matrices are successful. While exact numbers depend, of course, 
on the exact choice of cuts, tightening or loosening the cuts does not 
change one important feature:
Matrices of the form of Eq. \ref{Mbest} always do better than neutrino 
mass anarchy, especially if $r_{12}$, $\alpha$, $\beta$, $\gamma$ and 
$\delta$ are small numbers. 

This result is, of course, obtained by construction. Putting 
$\alpha=\beta=\gamma=\delta=0$ Eq. \ref{Mbest} is solved by a 
bi-maximal mixing matrix, which has $R=0$, $s_C=0$, $s_{atm}=1$ and 
$s_{\odot}=1$. This structure passes the cuts, and choosing the 
coefficients measuring departure from this case small enough, i.e. 
that they are only a slight perturbation, then simply yields the 
above numerical results. 

This example demonstrates that, even applying the cuts used in 
\cite{HMW}, structure 
in the neutrino mass matrix is obviously statistically preferred over 
neutrino mass anarchy. It should not be understood in the sense that 
Eq. \ref{Mbest} is necessarily the correct form of the neutrino mass 
matrix.

\end{document}